\documentclass[aip, rsi, amsmath, amssymb, reprint]{revtex4-1}

\usepackage{dcolumn}
\usepackage{bm}

\usepackage[utf8]{inputenc}
\usepackage[T1]{fontenc}
\usepackage{mathptmx}
\usepackage{amsmath, amsfonts, amssymb, graphicx}
\usepackage{textcomp}
\usepackage{gensymb}
\usepackage{caption}
\usepackage{float}
\usepackage{graphicx, subfig}
\usepackage[mathlines]{lineno}
\usepackage[colorlinks, linkcolor=blue, anchorcolor=blue, citecolor=cyan, urlcolor=cyan]{hyperref}

\newcommand{\zbldp}{\textrm{ZBL-DP}}
\newcommand{\zbl}{\textrm{ZBL}}
\newcommand{\deep}{\textrm{DP}}

\begin{document}

\preprint{AIP/123-QED}

\title{Deep learning inter-atomic potential model for accurate irradiation damage simulations}
\thanks{Hao Wang and Xun Guo contributed equally to this work.}

\author{Hao Wang}
 \affiliation{State Key Laboratory of Nuclear Physics and Technology, School of Physics, CAPT, HEDPS, and IFSA Collaborative Innovation Center of MoE College of Engineering, Peking University, Beijing 100871, P. R. China;}

\author{Xun Guo}
 \affiliation{State Key Laboratory of Nuclear Physics and Technology, School of Physics, CAPT, HEDPS, and IFSA Collaborative Innovation Center of MoE College of Engineering, Peking University, Beijing 100871, P. R. China;}

\author{Linfeng Zhang}
 \affiliation{Program in Applied and Computational Mathematics, Princeton University, Princeton, New Jersey 08544, USA;}

\author{Han Wang}
\email{wang\_han@iapcm.ac.cn}
\affiliation{Laboratory of Computational Physics, Institute of Applied Physics and Computational Mathematics, Beijing 100871, P. R. China;}

\author{Jianming Xue}
\email{jmxue@pku.edu.cn}
\affiliation{State Key Laboratory of Nuclear Physics and Technology, School of Physics, CAPT, HEDPS, and IFSA Collaborative Innovation Center of MoE College of Engineering, Peking University, Beijing 100871, P. R. China.}
 
\date{\today}

\begin{abstract}
We propose a hybrid scheme that interpolates smoothly the Ziegler-Biersack-Littmark (ZBL) screened nuclear repulsion potential with a newly developed deep learning potential energy model. The resulting DP-ZBL model can not only provide overall good performance on the predictions of near-equilibrium material properties but also capture the right physics when atoms are extremely close to each other, an event that frequently happens in computational simulations of irradiation damage events. We applied this scheme to the simulation of the irradiation damage processes in the face-centered-cubic aluminium system, and found better descriptions in terms of the defect formation energy, evolution of collision cascades, displacement threshold energy, and residual point defects, than the widely-adopted ZBL modified embedded atom method potentials and its variants. Our work provides a reliable and feasible scheme to accurately simulate the irradiation damage processes and opens up new opportunities to solve the predicament of lacking accurate potentials for enormous newly-discovered materials in the irradiation effect field.

\end{abstract}

\maketitle

With the rapid growth of computing science and computer performance, computational simulations, including molecular dynamics (MD)~\cite{Rn85,RN87,Rn88} and density functional theory (DFT) method~\cite{RN81, Rn83}, are becoming increasingly important to evaluate the properties of materials. However, the accuracy of empirically constructed atomic potential models for MD simulations are often in question, while the quantum mechanics approaches, such as DFT, are limited by the time and size scale of the simulated systems. 
Therefore, a solution that combines the advantages of both methods is needed.

Recently, machine learning (ML) methods have been used to solve this dilemma~\cite{RN17, RN27, RN28, RN29, RN30, RN31, RN42}. Several studies have demonstrated that the ML-based potential energy surface can reach the accuracy of DFT, with the cost comparable to classical empirical potentials~\cite{RN31, RN42, RN44, RN15}. Nevertheless, challenges have remained for ML-based methods to describe very short-distance interactions, e.g. those in the irradiation damage processes. 
In these processes, the distance between atoms can be very short, and the interactions can hardly be treated as quasi-static, wherein conventional DFT approaches may fail, so only the Ziegler-Biersack-Littmark (ZBL) screened nuclear repulsion potential~\cite{RN32} has been validated for a good description of the corresponding interactions.
In other words, in this case, energies and forces from DFT calculations may no longer be accurate training data for ML-based potentials.
Moreover, the magnitude of energies and forces is much larger than that in systems near equilibrium, which may pose additional difficulties for the training of ML-based potentials.
Therefore, it is necessary to develop a new scheme that is applicable for irradiation damage simulations while still remains the accurate predictions of material properties for both near-equilibrium state and short-distance interaction.

To solve this problem, 
we interpolate the ZBL potential into a deep learning model, so that short-distance collisions between atoms can be accurately described. 
In our previous studies, we have developed the Deep Potential (DP) scheme, an end-to-end symmetry preserving machine learning-based inter-atomic potential energy model, which can efficiently represent the properties of a wide variety of systems with the accuracy of ab-initio quantum mechanics models~\cite{RN28, RN17}. 
This ZBL-modified deep learning scheme (DP-ZBL), which can be seen as an improved and specialised version of the original DP model, makes it possible to accurately simulate the irradiation bombardment damage for materials. 
Here in this letter, we use face-centered-cubic (fcc) aluminium as the reference material, for which many irradiation experiments and collision simulations results have been reported~\cite{RN1, RN2, RN3, RN9, RN49, RN50, RN51}, to validate the feasibility and reliability of this method.

In the DP-ZBL model, we assume that the system under consideration is composed of $N$ atoms with coordinates denoted by
$\{ \bm R_i, \dots,  \bm R_N\}$. 
Similar to the original DP model~\cite{RN24, RN17},
the DP-ZBL potential assumes the system energy is decomposed into atomic contributions, i.e., 
\begin{eqnarray}
E^{\zbldp} = \sum_i E^\zbldp_i
\end{eqnarray}
with $i$ being the indexes of the atoms. 
The atomic contribution of atom $i$ is fully determined by the coordinates of atom $i$ and its near neighbors,

\begin{table*}
	\centering
	\caption{\label{table1} Equilibrium properties of Al: atomization energy $E_{\rm am}$, equilibrium lattice constant $a_{\rm 0}$, vacancy formation energy $E_{\rm vf}$, interstitial formation energy $E_{\rm if}$ for octahedral interstitial (oh) and tetrahedral interstitial (th), independent elastic constant $C_{\rm 11}$, $C_{\rm 12}$, and $C_{\rm 44}$, Bulk modulus $B_{V}$ (Voigt), shear modulus $G_{\rm V}$ (Voigt),stacking fault energy $\gamma_{\rm sf}$, twin stacking fault energy $\gamma_{\rm tsf}$, melt point $T_{\rm m}$, enthalpy of fusion $\Delta H_f$ and diffusion coefficient $D$ at $T$ = 1000 K.}
	
	\begin{ruledtabular}
		\begin{tabular}{ccccccc}
			Al\footnotemark[1]         &                  EXP.                   &           DFT            & DP-ZBL & DP\cite{RN14} & MEAM-ZBL & EAM-ZBL\footnotemark[2] \\ \hline
			$E_{\rm{am}}$ [eV/atom]       &            -3.49\cite{RN56}            &          -3.75           & -3.74  &       -3.65        &  -3.36   &  -3.39  \\
			$a_{\rm{0}}$ [\AA]         &            4.04\cite{RN57}             &           4.04           &  4.04  &        4.04        &   4.05   &  4.01   \\
			$E_{\rm{vf}}$ [eV]         &         0.66\cite{RN58, RN59}          &     0.67\cite{RN60}      &  0.73  &        0.79        &   0.67   &  1.14   \\
			$E_{\rm{if}}(oh)$ [eV]       &                   -                    &     2.91\cite{RN60}      &  2.57  &        2.45        &   3.12   &  -   \\
			$E_{\rm{if}}(th)$[eV]        &                   -                    &     3.23\cite{RN60}      &  3.23  &        3.12        &   3.83    &  -   \\
			$C_{\rm{11}}$ [GPa]         &            114.3\cite{RN61}            &          111.2           & 112.8  &       120.9        &  113.5   &  106.9  \\
			$C_{\rm{12}}$ [GPa]         &            61.9\cite{RN61}             &           61.4           &  57.6  &        59.6        &   61.6   &  81.5   \\
			$C_{\rm{44}}$ [GPa]         &            31.6\cite{RN61}             &           36.8           &  41.2  &        40.4        &   45.4   &  44.2   \\
			$B_{\rm{V}}$ [GPa]         &            79.4\cite{RN61}             &           78.0           &  76.0  &        80.1        &   78.9   &  90.0   \\
			$G_{\rm{V}}$ [GPa]         &            29.4\cite{RN61}             &           32.1           &  35.8  &        36.5        &   37.6   &  31.6   \\
			$\gamma_{\rm{sf}}{\rm [J/m^{2}]}$  & 0.11-0.21\cite{RN62, RN63, RN64, RN65} &     0.142\cite{RN66}     &  0.070 &       0.132        &  0.184   &   -   \\
			$\gamma_{\rm{tsf}}{\rm [J/m^{2}]}$ &                   -                    &     0.135\cite{RN66}     &  0.075 &       0.130        &  0.184   &   -   \\
			$T_{\rm{m}}$ [K]          &             935\cite{RN67}             & 950($\pm 50$)\cite{RN68} &  885   &         918        &   950    &   1050   \\
			$\Delta H_{\rm{f}}$ [KJ/mol]    &       10.7($\pm$ 0.2)\cite{RN69}       &            -             &  9.3   &        10.2        &   11.5   &   8.8   \\
			$D{ \rm [10^{-9}m^2/s]}$      &           7.2-7.9\cite{RN70}           &            -             &  6.8   &        7.1         &   4.9    &   6.8
		\end{tabular}
	\end{ruledtabular}
	\footnotetext[1]{The results above, unless specified with a reference, are computed by the authors.} 
	\footnotetext[2]{The interstitial and stacking fault configurations were unstable upon relaxation with the EAM-ZBL potential, so their formation energies are not reported here.} 
\end{table*}
 
\begin{align}
 E^\zbldp_i = E^\zbldp_{s(i)} (\bm R_i, \{\bm R_j \vert j\in \mathcal N_{R_c}(i)\}) 
\end{align}
where $s(i)$ denotes the chemical species of atom $i$, 
and $\mathcal N_{R_c}(i)$ denotes the set of near neighbors within cut-off radius $R_c$, 
i.e.~$\mathcal N_{R_c}(i) = \{j\vert R_{ij} = \vert\bm R_{ij}\vert \leq R_c\}$.
The atom contribution of DP-ZBL is the interpolation of
the ZBL screened nuclear repulsion potential $E^\zbl_i$ and the standard deep potential $E^\deep_i$
\begin{align}
 E^\zbldp_i= w_i E^\zbl_i + (1 - w_i) E^{\deep}_i, 
\end{align}
where $w_i$ is the scale of ZBL potential that smoothly changes from 1 to 0 as the distance between atom $i$ and its \emph{nearest} neighbor goes from 0 to a threshold value.
To be more specific, the scale $w_i$ is defined as
\begin{align}
\label{fun:4}
 w_i = \left\{
 \begin{aligned}
  &1 && \sigma_i < R_a, \\
  &-6 u_i^5 + 15 u_i^4 - 10 u_i^3 + 1 && R_a \leq \sigma_i < R_b, \\
  &0 && \sigma_i \geq R_b,
 \end{aligned}
 \right.
\end{align}
with $u_i$ being the short-hand notation defined by
\begin{align}
 u_i = \dfrac{\sigma_i - R_a}{R_b - R_a},
\end{align}
and $[R_a, R_b)$ denoting the range in which the ZBL potential and the deep potential are interpolated.
It is noted that the switch function $-6 u_i^5 + 15 u_i^4 - 10 u_i^3 + 1$ is continuous at 0 and 1 up to the second order derivative.
The symbol $\sigma_i$ denotes the smooth-minimal distance of atom $i$'s near neighbors, which is defined by
\begin{align}
 \sigma_i =
 \dfrac
 {\sum_{j\in\mathcal N_{R_c}(i)} R_{ij} e^{-R_{ij} / \alpha} }
 {\sum_{j\in\mathcal N_{R_c}(i)} e^{-R_{ij} / \alpha} },
\end{align}
with $\alpha$ being a tunable scale of the distances between atoms. In the current work, we fix the scale to $\alpha$ = 0.1~\AA.

\begin{figure}
	\includegraphics[width=0.45\textwidth]{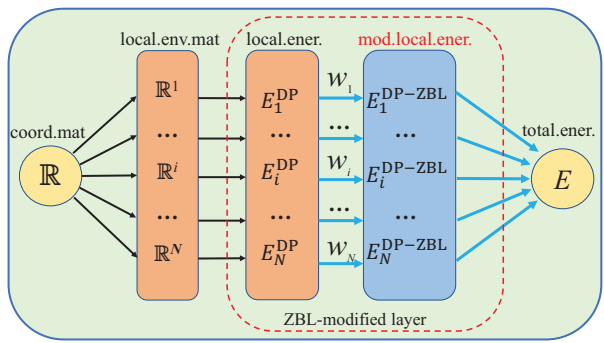}
	\caption{\label{fig:1} Schematic plot of the DP-ZBL model. In the mapping from the coordinate matrix $\mathbb{R}$ to the potential energy $E$, $\mathbb{R}$ is firstly transformed to local environment matrices $\{\mathbb{R}^i\}_{i=1}^{N}$. Then each $\mathbb{R}^i$ is mapped, through a sub-network to a local "atomic" energy $E_i^{\rm DP}$ as well as the original DP model. Then a ZBL-modified layer is added to better reproduce the strong repulsion at short inter-atomic distances, through the smooth interpolation of the ZBL screened nuclear repulsion potential $E_i^{\rm ZBL}$ and the standard deep potential $E_{i}^{\rm DP}$. Finally, $E_{total}=\sum_i E_i^{\rm DP-ZBL}$.}
\end{figure}

As the schematic diagram is shown in FIG. \ref{fig:1}, a ZBL-modified layer is added to better describe the strong repulsion at short inter-atomic distances in the DP-ZBL model, through the smooth switch function $w_i$ in Eq. \ref{fun:4}. Then the DP-ZBL model neural network was trained with the same dataset generated by the deep potential generator (DP-GEN) in Ref. [\onlinecite{RN14}], a scheme employing the idea of active learning~\cite{settles2012active} and reinforced dynamics~\cite{RN22}. Note that this dataset contains a vast range of configurations explored and labelled during the active-learning process, which ensures that the DP-ZBL model can be trained with enough possible configurations with high accuracy. Apart from the interpolation with the ZBL potential, the cut-off radius adopted by the current work is 6 \AA, and the total training steps are 640, 000. These differences in training will not lead to a significant difference in the accuracy.
We have also tested several switching ranges ($R_{a}$, $R_{b}$) to generate the DP-ZBL potentials. The best one (1.2\AA, 2.0\AA) was selected for collision cascades simulations that are presented below.

\captionsetup[figure]{labelfont=bf}
\captionsetup[table]{labelfont=bf}

In order to evaluate the behavior of the DP-ZBL potential on irradiation effects, two classical potentials have also been employed for comparison, including the ZBL joined embedded atom method (EAM) potential (EAM-ZBL)\cite{RN48} and the state-of-the-art modified EAM (MEAM) potential~\cite{RN5} with self-implemented ZBL (MEAM-ZBL), which are widely used in the previous irradiation simulations and give satisfying results~\cite{RN71, RN73, RN74}. In this work, we used the DeePMD-kit\cite{RN20} for training the DP-ZBL potential, LAMMPS\cite{RN40} for molecular dynamic simulations, VASP\cite{RN34, RN35, RN36} for ab-initio calculations, and OVITO\cite{RN39} for the defect identification.

First, we have calculated some material properties using the three potentials and the DFT (see Methods in the supplementary materials), as summarized in TABLE \ref{table1}. It is no wonder that the MEAM-ZBL potential provides nearly the same vacancy formation energies ($E_{\rm vf}$) as the results of experiments because these basic solid state properties have been used to tune the parameters of the MEAM potentials. Besides that, the DP-ZBL potential gives reliable results in all these considered properties. 
These results demonstrate that the DP-ZBL potential can still provide accurate predictions about the material properties near the equilibrium state with the accuracy comparable to the DFT calculations, which also implies that the smoothly joined ZBL potential which dominates the interatomic interactions below 1.2 \AA \ would not influence the accuracy of the original DP model.

\begin{figure}[hb]
	\includegraphics[width=0.45\textwidth]{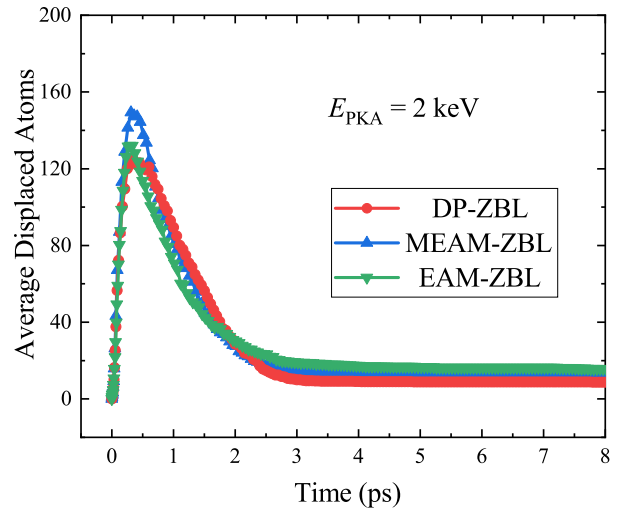}
	\caption{\label{fig:2} The number of average displaced atoms $N_{\rm d}$ during the evolution of the collision cascade caused by a 2 keV PKA. Each point is the average of 10 independent 2 keV cascade simulations.}
\end{figure}

Next, we did collision cascade simulations by using these three potentials. Collision cascades are the feature phenomena in irradiation effects~\cite{RN75, RN76, RN77}. When the energetic particles including protons, neutrons, electrons, and ions inject the target material, it will transfer energy to the target atoms. If the transferred energy is higher than the displacement threshold energy ($E_{\rm d}$) of target atoms, they will displace from the original lattice sites. If these primary knock-on atoms (PKAs) still have enough energy, they can knock out other target atoms subsequently and so on. Thus a large number of atoms are displaced from their original lattice sites, which is called the collision cascade. However, as the cascade begins to thermally equilibrate with its surrounding environment, most of the displaced atoms regain position in the perfect lattice structure\cite{RN79, RN80}, as illustrated in FIG. \ref{fig:2}.

It can be observed in FIG. \ref{fig:2} that all the three potentials exhibit a similar trend of displaced atoms during the evolution. The number of displaced atoms increased sharply within 1 ps and reached a peak at 0.3 $\sim$ 0.4 ps. Then it decreased monotonically because of the recombination process, and only a few defects remained. The evolution of displaced atoms generated by PKA at other energies was also illustrated in the Supplementary Materials (SM). It can be concluded from FIG. S5 that the peak value of displaced atoms increased with the increasing PKA energy, but the peak of MEAM-ZBL model is significantly higher than other models when the PKA energy is larger than 2~keV. Though this transient process can hardly be examined by experiment or other models, which means we cannot give a reliable estimation which one is more accurate, we can still conclude that the collision cascade evolutions provided by DP-ZBL potential do not significantly deviate from the results obtained other existing methods.

In fact, the number of residual point defects is even more important than the peak value during evolution, for a broad range of fundamental science and applied engineering applications. To quantify the numbers of point defects caused by a single PKA, Norgett $et~al.$ have proposed the Norgett-Robinson-Torrens (NRT) model, based on the binary collision approximation method, to evaluate the bombardment damage~\cite{RN45, RN46}. However, it has been recognized for several decades that the NRT model overestimates near 3 times the number of stable defects in pure metals after energetic cascades~\cite{RN52, RN53, RN54}. Therefore, we calculated the residual point defects by the NRT model and used one third of it as a benchmark to evaluate our cascade simulation results, and a brief introduction of NRT model was also introduced in the SM.




\begin{figure}
	\includegraphics[width=0.45\textwidth]{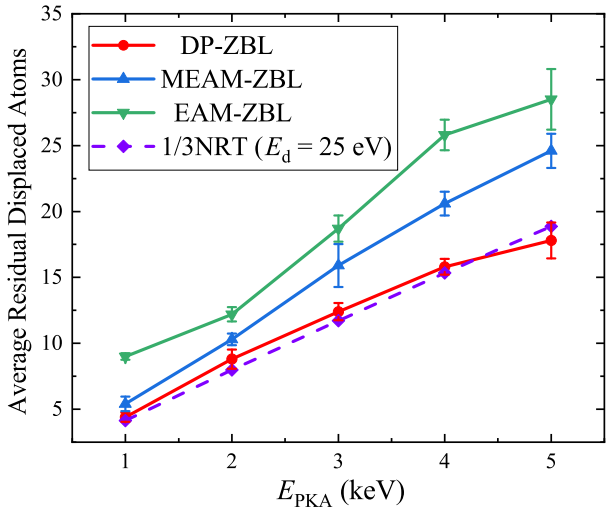}
	\caption{\label{fig:3} The residual point defects after 50 ps relaxation and the corresponding 1/3 NRT model results. Each point is the average of 10 independent cascade simulations, and the errors are given in the standard error of the mean.}
\end{figure}

As shown in FIG. \ref{fig:3}, the residual displaced atoms calculated by the three potentials all exhibit similar near-linear trends with the value of initial PKA energy, which is consistent with the NRT model according to Eq. S2. Note that the slope of the fitting line is inversely proportional to the displacement energy $E_{\rm d}$, which is usually defined as the minimum amount of energy transferred to a lattice atom to make it displace from original stable site~\cite{RN26}. Other widely used models, such as the Kinchin-Pease (KP) model~\cite{RN47} and the athermal recombination corrected DPA (arc-DPa) model~\cite{RN10}, typically take $E_{\rm d}$ as a key parameter to quantify the amount of displacement damage generated by energetic particles inject in materials. So we also made a comparison of average $E_{\rm d}$ value in all the possible directions calculated by these three models, as shown in TABLE~\ref{table2} and FIG. S4. 

\begin{table}[b]
	\caption{\label{table2}The average displacement threshold energy ($E_{\rm d}$) of fcc Al.}
	\begin{ruledtabular}
		\begin{tabular}{ccccc}
			fcc Al & Recommended\cite{RN55} & DP-ZBL & MEAM-ZBL & EAM-ZBL \\
			\hline
			$E_{\rm d}\ ({\rm eV})$ & 25.0 & 26.54 &22.67 &16.73\\
		\end{tabular}
	\end{ruledtabular}
\end{table}

According to our calculations, $E_{\rm d}$'s provided by the DP-ZBL and MEAM-ZBL potentials were quite close to the recommended value (25 eV for fcc Al) \cite{RN55, RN26}, while the result of EAM-ZBL potential has a relatively large deviation. Furthermore, the DP-ZBL potential can also give the best prediction of residual displaced atoms in the three examined potentials, if we took the 1/3 NRT results as a reference. Therefore, the simulation results of DP-ZBL potential are consistent with most of the existing theoretical models in the field of low-energy ion irradiation damage.

Meanwhile, the displacement energy of three potentials are in the order of $E_{\rm d}$(DP-ZBL) > $E_{\rm d}$(MEAM-ZBL) > $E_{\rm d}$(EAM-ZBL), so the corresponding residual point defects values should be $N_{\rm d}$(EAM-ZBL) > $N_{\rm d}$(MEAM-ZBL) > $N_{\rm d}$(DP-ZBL), which is completely in accordance with the results in FIG.~\ref{fig:3}. Besides, it is worth noting that the DP-ZBL model usually produces smaller $N_{\rm d}$ than the classical EAM-ZBL and MEAM-ZBL potentials, which may be caused by the fact that the DFT data sets used to train the DP-ZBL model have considered the energy difference of configurations far from the equilibrium state, while the traditional EAM-ZBL and MEAM-ZBL potentials are simply constructed by fitting the material properties near the equilibrium state. From this point of view, the DP-ZBL model should provide a better description for the irradiation damage events than traditional analytical potentials. 

In conclusion, we have proposed the DP-ZBL scheme by smoothly interpolating the accurate repulsive pair potential (ZBL) into the DP model. The DP-ZBL potential generated in this way can not only give accurate results regarding the material properties near equilibrium states but also be sufficient to describe the atomic collision cascades during the irradiation damage processes. This method can minimize the impact of subjective factors on potentials during their establishments, and provide higher agreement with experimental or DFT results compared with other widely used classical potentials. Moreover, due to the applicability of the DP model to a wide range of materials, see, e.g. Ref.[\onlinecite{RN17}], this newly proposed method may be used in the irradiation effect studies of new advanced materials, such as high entropy alloys (HEAs), layered transition metal ternary nitrides and carbides ($\rm{M_{n+1}AX_n}$ phases), and two-dimensional (2D) materials, for whom suitable classical potentials are still lacking. We hope that with the development and improvement of the DP-ZBL potential database and corresponding computational algorithm, the predicament of accurate potentials lacking in the irradiation effect field could be better solved.

\begin{acknowledgments}
This work is supported by National Natural Science Foundation of China (Grant No. 11705010 and 11871110), and the National Key Research and Development Program of China (Grants No. 2016YFB0201200 and 2016YFB0201203). We are grateful for computing resource provided by Weiming No. 1 and Life Science No. 1 High Performance Computing Platform at Peking University, the Terascale Infrastructure for Groundbreaking Research in Science and Engineering (TIGRESS) High Performance Computing Center and Visualization Laboratory at Princeton University, as well as TianHe-1(A) at National Supercomputer Center in Tianjin.
\end{acknowledgments}

\end{document}


\preprint{AIP/123-QED}
\title[Supplementary Materials]{Deep learning inter-atomic potential model for accurate irradiation damage simulations}
\thanks{Hao Wang and Xun Guo contributed equally to this work.}

\author{Hao Wang}
 \affiliation{State Key Laboratory of Nuclear Physics and Technology, School of Physics, CAPT, HEDPS, and IFSA Collaborative Innovation Center of MoE College of Engineering, Peking University, Beijing, People's Republic of China}

\author{Guo Xun}%
 \affiliation{State Key Laboratory of Nuclear Physics and Technology, School of Physics, CAPT, HEDPS, and IFSA Collaborative Innovation Center of MoE College of Engineering, Peking University, Beijing, People's Republic of China
}%

\author{Linfeng Zhang}
 \affiliation{Program in Applied and Computational Mathematics, Princeton University, Princeton, New Jersey 08544, USA}

\author{Han Wang}
\email{wang_han@iapcm.ac.cn}
 \affiliation{Laboratory of Computational Physics, Institute of Applied Physics and Computational Mathematics, Huayuan Road 6, Beijing 100088, People’s Republic of China}

\author{Jianming Xue}
\email{jmxue@pku.edu.cn}
 \affiliation{State Key Laboratory of Nuclear Physics and Technology, School of Physics, CAPT, HEDPS, and IFSA Collaborative Innovation Center of MoE College of Engineering, Peking University, Beijing, People's Republic of China}

\date{\today}

\begin{abstract}
	We propose a hybrid scheme that interpolates smoothly the Ziegler-Biersack-Littmark (ZBL) screened nuclear repulsion potential with a newly developed deep learning potential energy model. The resulting DP-ZBL model can not only provide overall good performance on the predictions of near-equilibrium material properties but also capture the right physics when atoms are extremely close to each other, an event that frequently happens in computational simulations of irradiation damage events. We applied this scheme to the simulation of the irradiation damage processes in the face-centered-cubic aluminium system, and found better descriptions in terms of the defect formation energy, evolution of collision cascades, displacement threshold energy, and residual point defects, than the widely-adopted ZBL modified embedded atom method potentials and its variants. Our work provides a reliable and feasible scheme to accurately simulate the irradiation damage processes and opens up new opportunities to solve the predicament of lacking accurate potentials for enormous newly-discovered materials in the irradiation effect field.
	
\end{abstract}

\maketitle

\appendix

\captionsetup[figure]{labelfont=bf}
\captionsetup[table]{labelfont=bf}
\renewcommand\thefigure{ S\arabic{figure}}

\section{Methods}
\subsection{Molecular Dynamic Simulations}
All our molecular dynamic simulations are carried out using the LAMMPS software.\cite{RN40} 

\textbf{Pair potential and force calculation :} In order to quickly evaluate whether different potentials are sufficient to reproduce the strong repulsive interactions between atoms nearer than 1 \AA, we firstly calculate the pair potential and force between one pair atoms, by fixing one atom unmoved and gradually moving another atom. In practice, the initial distance of the pair atoms is 5 \AA, \ then we decrease their distance 0.002 \AA \ every step and record the corresponding pair energy and force.

\begin{figure}[H]
\centering
\subfloat[]{
\includegraphics[width=0.6\textwidth]{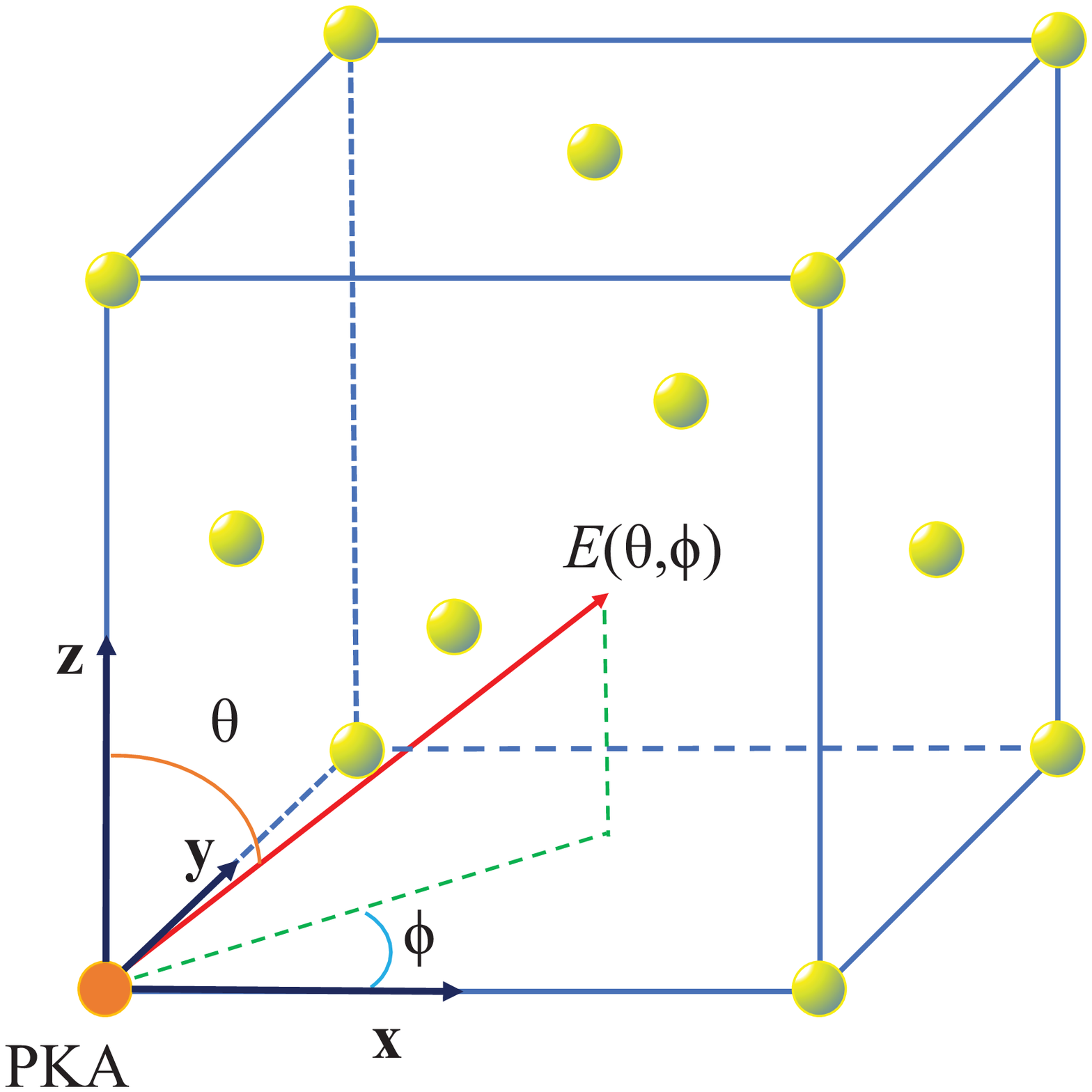}}
\caption{The schematic plot of calculating the $E_{\rm d}(\theta, \phi)$.}
\label{fig:s1}
\end{figure}

\textbf{Displacement threshold energy :} The displacement threshold energy $E_{\rm d}$ is defined as a spherical average of potential barrier surrounding the equilibrium lattice site. As shown in the FIG. \ref{fig:s1}, in the 500 atoms system, by given the chosen atom (PKA) a velocity $(v \sin\theta \cos\phi, v\sin\theta \sin\phi$, $v\cos\theta)$ $(v=\sqrt{2E_{\rm PKA}/{m}})$ to find the minimum energy needed to form a stable Frankel pair, we have calculated the $E_{\rm d}(\theta, \phi)$ every $5^\circ$ ($0^\circ \leq \theta \leq 90^\circ$, $0^\circ \leq \phi \leq 90^\circ$), for example ($0^\circ, 0^\circ$), ($0^\circ, 5^\circ)$,...,($90^\circ, 90^\circ$). The corresponding contour plots are provided in FIG. S4(a-c) below.

\textbf{Defect formation energy :} The formula widely-used to calculate the defect formation energy in the electric neutral system is given below.
\begin{equation*}
\Delta E_{i} = E_{i}({\rm defect}) - E({\rm supercell}) - \sum_{s=1}^{N_{species}} n_{s}^{i}\mu_{s} \eqno{\rm (S1)}
\end{equation*}
Where $E_{i}({\rm defect})$ is the total energy of a supercell with the defect; $E({\rm supercell})$ is the energy of the perfect bulk supercell; $n_{s}^{i}$ is the number of atoms of type $s$ that were added ($n_{s}^{i}$ > 0) or removed ($n_{s}^{i}$) to create the defect; $\mu_s$ is the chemical potential of atomic specie $s$; and the sum is taken over all elemental species, $N_{\rm species}$. In order to get a accurate defect formation energy, a large super cell which contains 4000 Al atoms was used in our MD simulations. 

\textbf{Collision cascade simulations:}
For PKAs' energy in 1 $\sim$ 5 keV, a system containing 600000 atoms was used (nearly $16 \times 16 \times 20 \ nm^{3}$ large). All the systems were relaxed under 300 k NVT ensemble using the Nose-Hoover thermostat\cite{RN41} for 10 ps. Then a random atom was selected as the PKA, and then given a velocity along the Z direction. The collision cascade simulations were under the NVE ensemble for 50 ps to get a stable result of residual point defects. A variable timestep was adopted to insure that the maxmium distance of one atom during one timestep is below 0.05 \AA. Besides, the Weigner-Seitz method implanted in OVITO\cite{RN39} is employed to identify the point defects including vacancies (VACs) and self-interstitials (SIAs) during cascade processes.


\subsection{DFT calculations}
In our work, the first-principles calculations were performed using the projected augmented wave (PAW) method\cite{RN33} as implemented in Vienna ab initio simulation package (VASP).\cite{RN34,RN35,RN36} The exchange-correction interaction was treated by the Perdew-Burke-Ernzerhof (GGA-PBE) exchange correlation functional.\cite{RN37} We have employed a plane-wave basis set with an energy cutoff of 600 eV and Monkhorst-Pack $k$-point mesh\cite{RN38} of 3 $\times$ 3 $\times$ 3 to calculate the corresponding defect formation energy in a 500 atoms super-cell.
\subsection{NRT model}
In 1975, Norgett $et~al.$ proposed the simple NRT formula\cite{RN46} to calculate the number of Frenkel pairs $N_{\rm d}$ generated by a primary knock-on atom of initial kinetic energy $E$:

\begin{equation*}
 N_{\rm d}(T_{\rm d}) = \left[
 \begin{aligned}
  &0, &     &T_{\rm d}< E_{\rm d} \\
  &1, &   E_{\rm d} &\leq T_{\rm d}< \dfrac{2E_{\rm d}}{0.8} \\
  &\dfrac{0.8T_{\rm d}}{2E_{\rm d}}, &  &\dfrac{2E_{\rm d}}{0.8}\leq T_{\rm d} 
 \end{aligned}
 \right] 
 \eqno{\rm (S2)}
\end{equation*}

where $T_{\rm d}$ is the energy available to generate atomic displacements by elastic collisions and the $E_{\rm d}$ is the displacement threshold energy. The elastic energy loss is calculated according to the method of Lindhard et al. using a numerical approximation to the universal function $g(\epsilon)$:
\begin{equation*}
T_{\rm d} = \dfrac{E}{[1 + {\rm k} g(\epsilon)]}, \eqno{\rm (S3)}
\end{equation*}
\begin{equation*}
g(\epsilon)= 3.4008 \epsilon^{1/6} + 0.40244 \epsilon^{3/4} + \epsilon, \eqno{\rm (S4)}
\end{equation*}
\begin{equation*}
k = 0.1337 Z_{1}^{1/6}(Z_1/A_1)^{1/2}, \eqno{\rm (S5)}
\end{equation*}
\begin{equation*}
\epsilon = [A_2E/(A_1 + A_2)] [a/(Z_1Z_2e^2)] \eqno{\rm (S6)}
\end{equation*}
\begin{equation*}
a = (9\pi^2/128)^{1/3}a_{0}[Z_1^{2/3} + Z_2^{2/3}]^{-1/2} \eqno{\rm (S7)}
\end{equation*}
where $a_{0}$ is the Bohr radius, $e$ the electronic charge, $Z_1$ and $Z_2$ are the atomic numbers of the projectile and target and $A_1$ and $A_2$ are the mass numbers of the two atoms.

The NRT model is based on the more than 40-year-old binary collision computer simulations of ion collision in solids and overestimates nearly three times the number of stable defects caused by energetic cascades in pure metals.\cite{RN10,RN52,Rn53,RN54} So in this letter, we take 1/3 the NRT results as a benchmark.

\section{Some testing results}
\subsection{Pair potential and force using different potentials}
We have tested two empirical potentials and two machine-learning potentials of the element aluminum (Al) to evaluate whether they are sufficient to describe the interactions between the pair atoms below 1.0 \AA,\ compared with the accurate ZBL repulsive pair potential.\cite{RN32} As shown in FIG. 1, all the four potentials including eam,\cite{RN48} eam\_fs,\cite{RN49} agni(2015PRB)\cite{RN50} and agni(2017JPCC)\cite{RN51} are not sufficient to reproduce the strong repulsive interatomic interactions below 1 \AA. \ In detail, the agni(2015PRB) and agni(2017JPCC) potentials nearly donot have the repulsive potential, while the eam potential underestimates the repulsive interactions. Note that the eam\_fs potential gives almost the same pair potential and force as the ZBL potential in the range (0.5 \AA, \ 2.0 \AA), however this potential has a discontinuity point near 0.5 \AA \ and then overestimates the repulsive interaction 8 orders of magnitude than the ZBL potential below 0.5 \AA. In conclusion, without the accurate description of the pair potential and force below 1 \AA, \ these potentials cannot be used in irradiation damage simulations.

\begin{figure}[H]
\centering
\subfloat[]{
\includegraphics[width=0.47\textwidth]{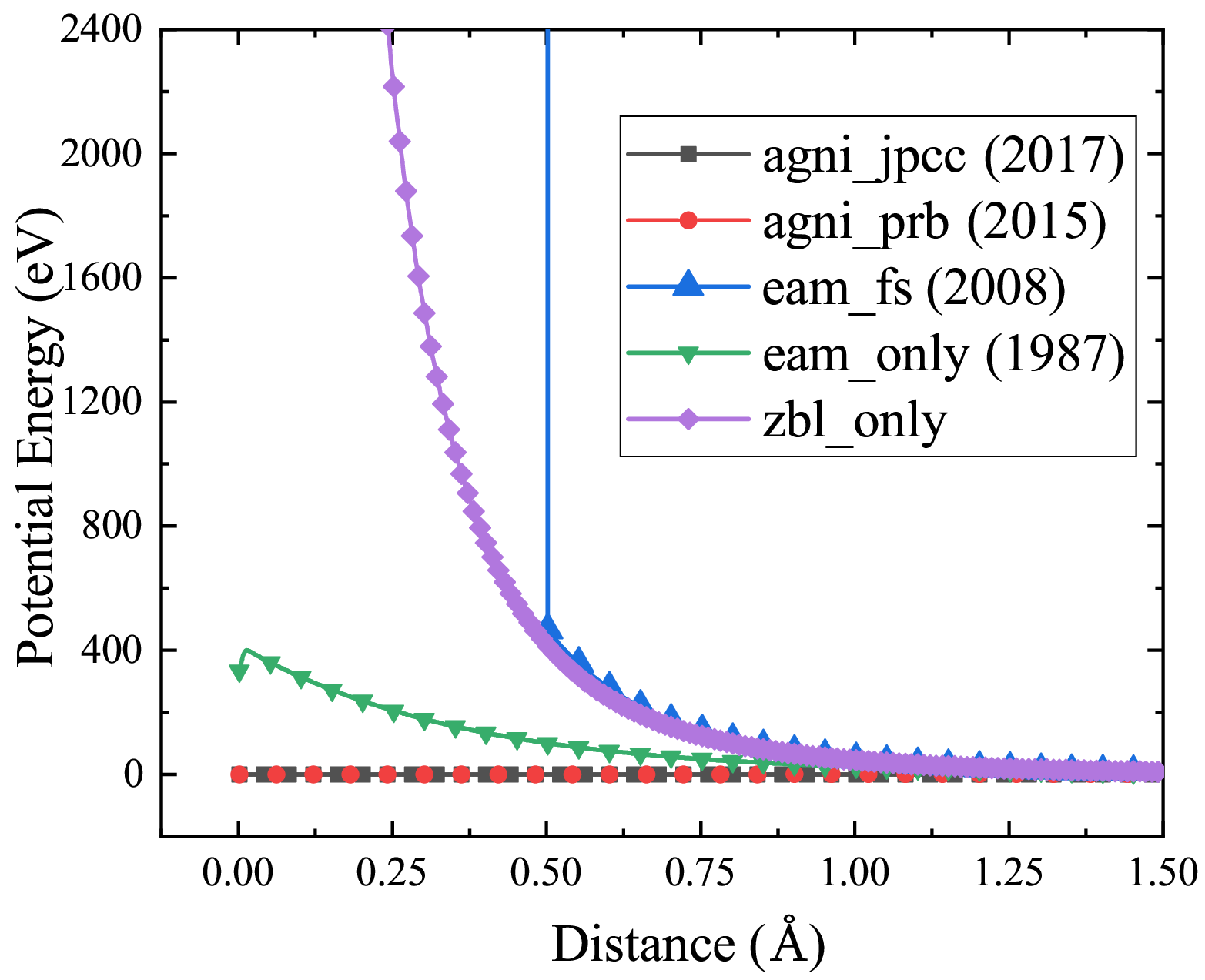}}
\quad
\subfloat[]{
\includegraphics[width=0.49\textwidth]{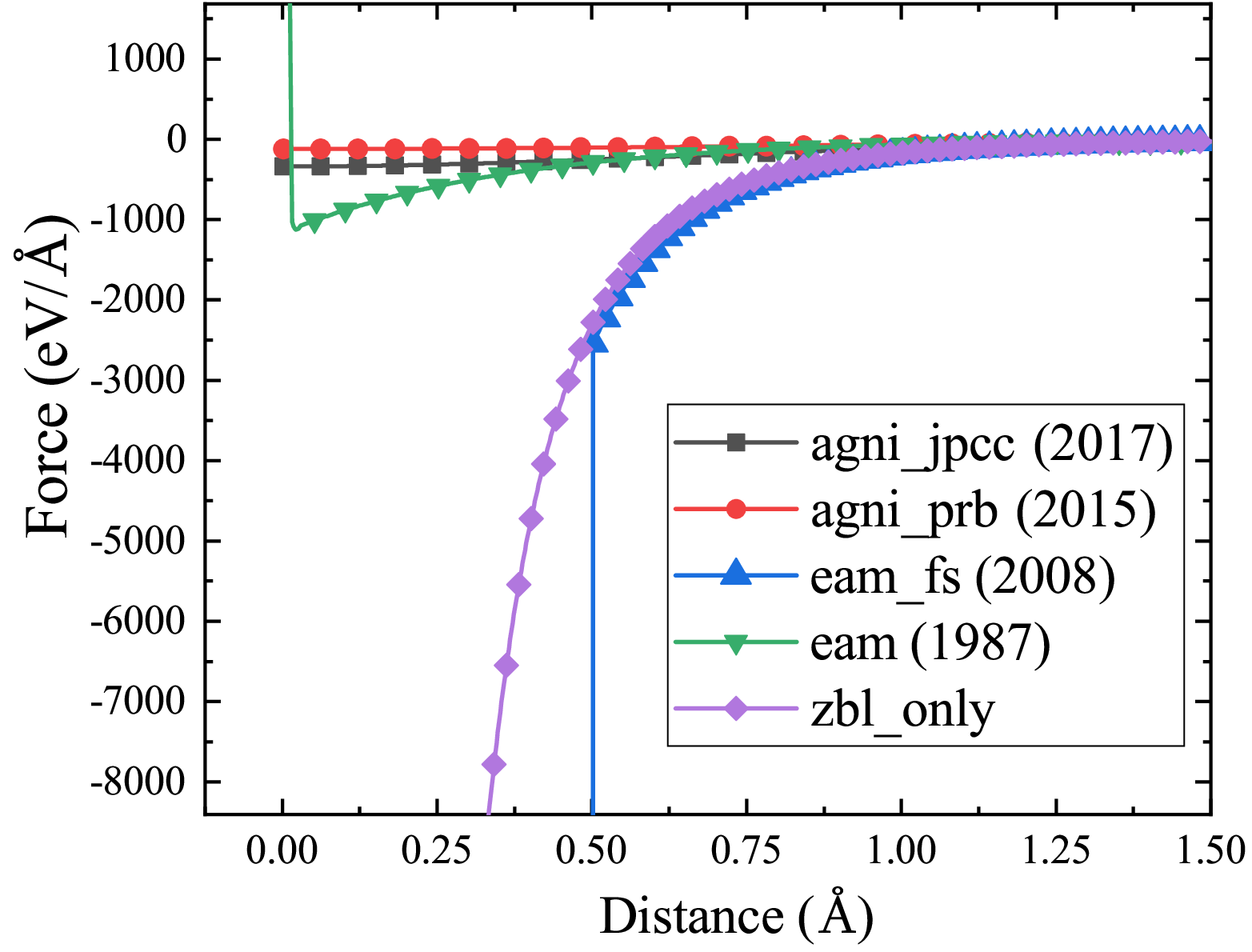}}
\caption{The pair potential (a) and force (b) calculated using two classical potentials (eam and eam.fs) and two machine learning potentials (agni\_jpcc and agni\_prb). Compared the ZBL repulsive pair potential, all the four potentials cannot give an accurate description about the interatomic interactions below 1 \AA.}
\end{figure}

\newpage
Then we have calculated the corresponding pair potential and force of the specially constructed DP-ZBL deep-learning potential, ZBL modified 
eam (EAM-ZBL) and the state of the art classical potential (MEAM-ZBL) with self-implemented ZBL potential in LAMMPS. As the three potentials all have smoothly transferred to the ZBL repulsive pair potential at short interatomic distances shown in FIG. S3, all of them can be used to describe the inter-atomic interactions between atoms during the cascade events. Besides that, it's noted that the DP-ZBL potential reproduces exactly the same pair potential and force as the ZBL potential below 1.2 \AA \ as we expected. However, the MEAM-ZBL potential overestimates the repulsive interaction below 1 \AA \ more or less, compared with the accurate ZBL potential. From this point of view, the DP-ZBL potential should have better performance on the collision cascade simulations even compared with the state-of-art classical potential MEAM-ZBL.

\begin{figure}[H]
\centering
\subfloat[]{
\includegraphics[width=0.46\textwidth]{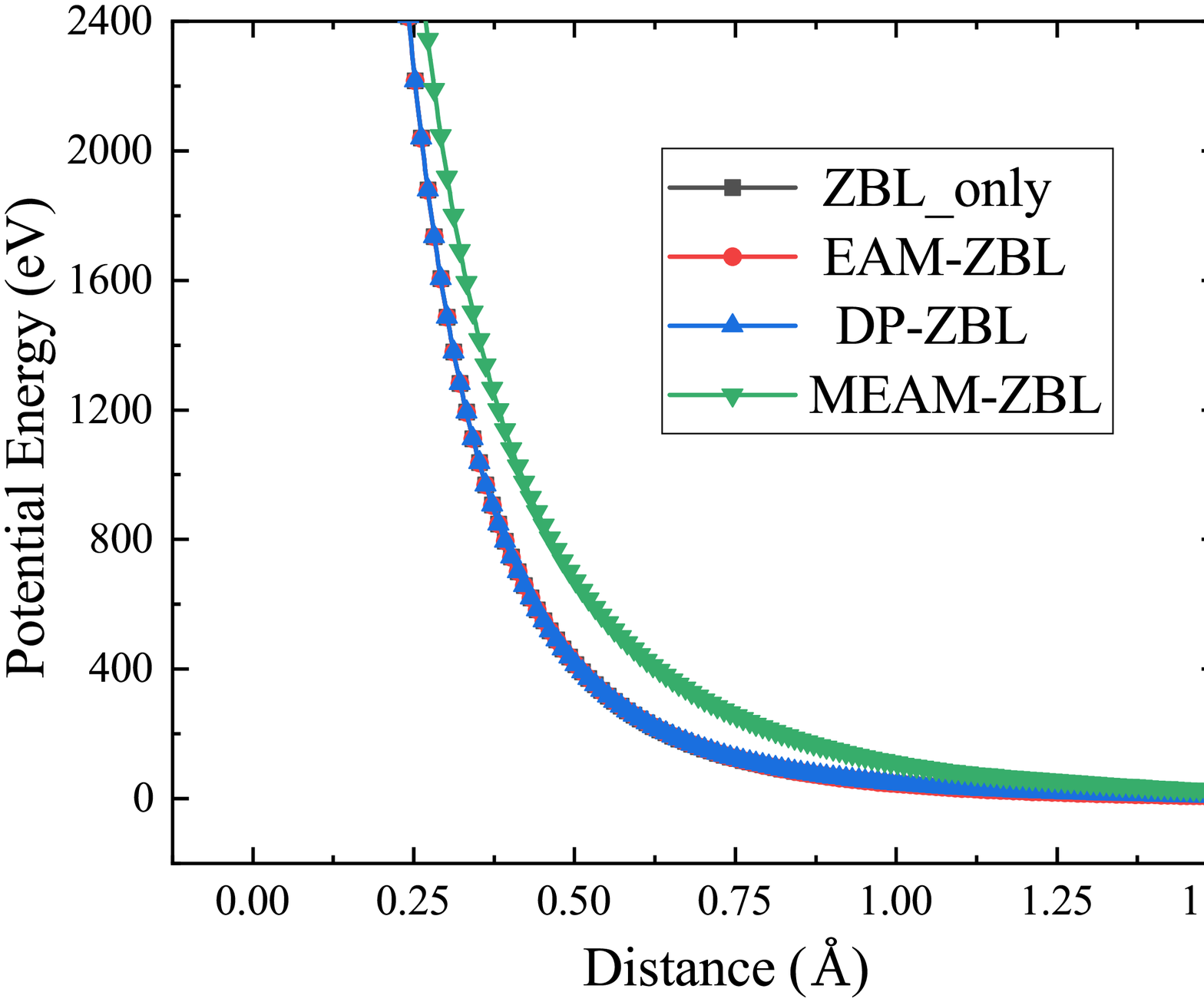}}
\quad
\subfloat[]{
\includegraphics[width=0.48\textwidth]{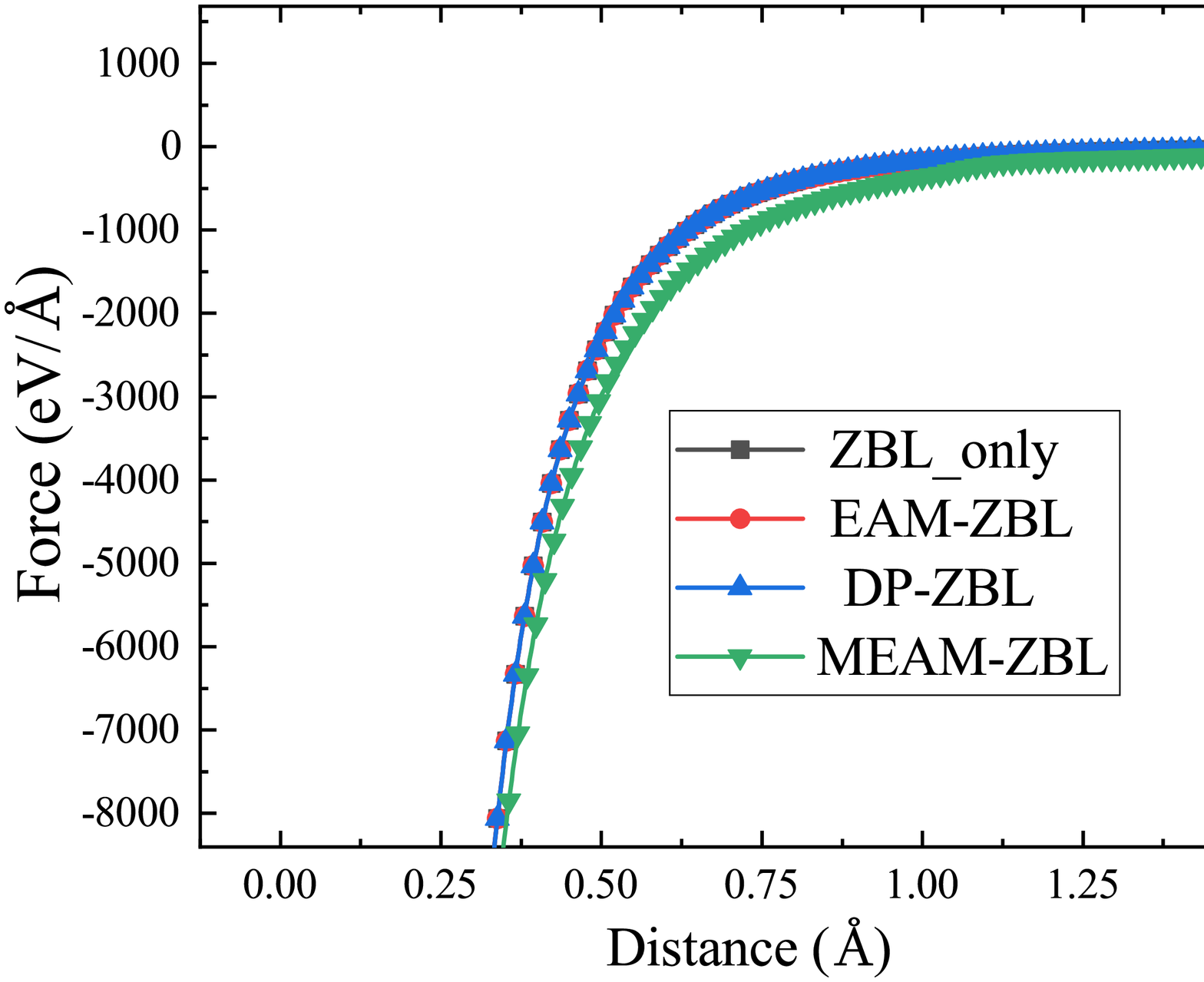}}
\caption{The pair potential (a) and force (b) calculated using the DP-ZBL, EAM-ZBL and one state-of-art classical potentials MEAM-ZBL. As all these potentials are smoothly transformed into the ZBL repulsive potential at short distance, they are sufficient to describe the interatomic interactions during the cascade simulations.}
\label{fig:s3}
\end{figure}

\subsection{The displacement threshold energy}
Based on the methods above, we have calculated the average displacement threshold energy $E_{\rm d}$ of three potentials including the DP-ZBL, MEAM-ZBL and EAM-ZBL potentials. The corresponding $E_{\rm d}(\theta, \phi)$ are provided in FIG. S4(a-c) below, which demonstrates that all the three potentials show similar dependence of $E_{\rm d}$ on the PKA initial direction ($\theta, \phi$). As the recommended $E_{\rm d}$ value in FCC Al in 25 eV, both the DP-ZBL, and MEAM-ZBL potentials can give a reliable $E_{\rm d}$, while the EAM-ZBL potential underestimates it. 

\begin{figure}
\centering
\subfloat[]{
\includegraphics[width=0.5\textwidth]{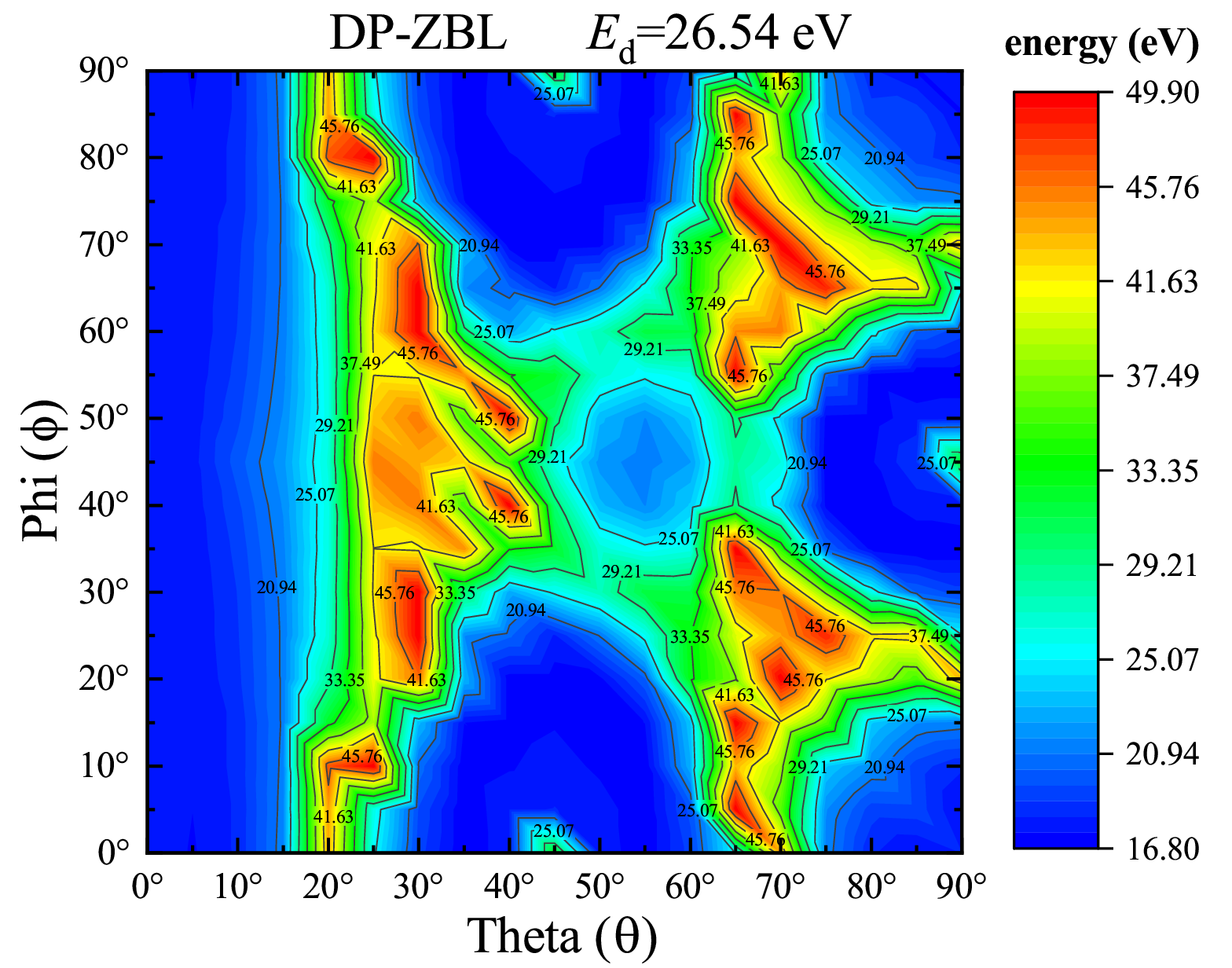}}
\quad
\subfloat[]{
\includegraphics[width=0.5\textwidth]{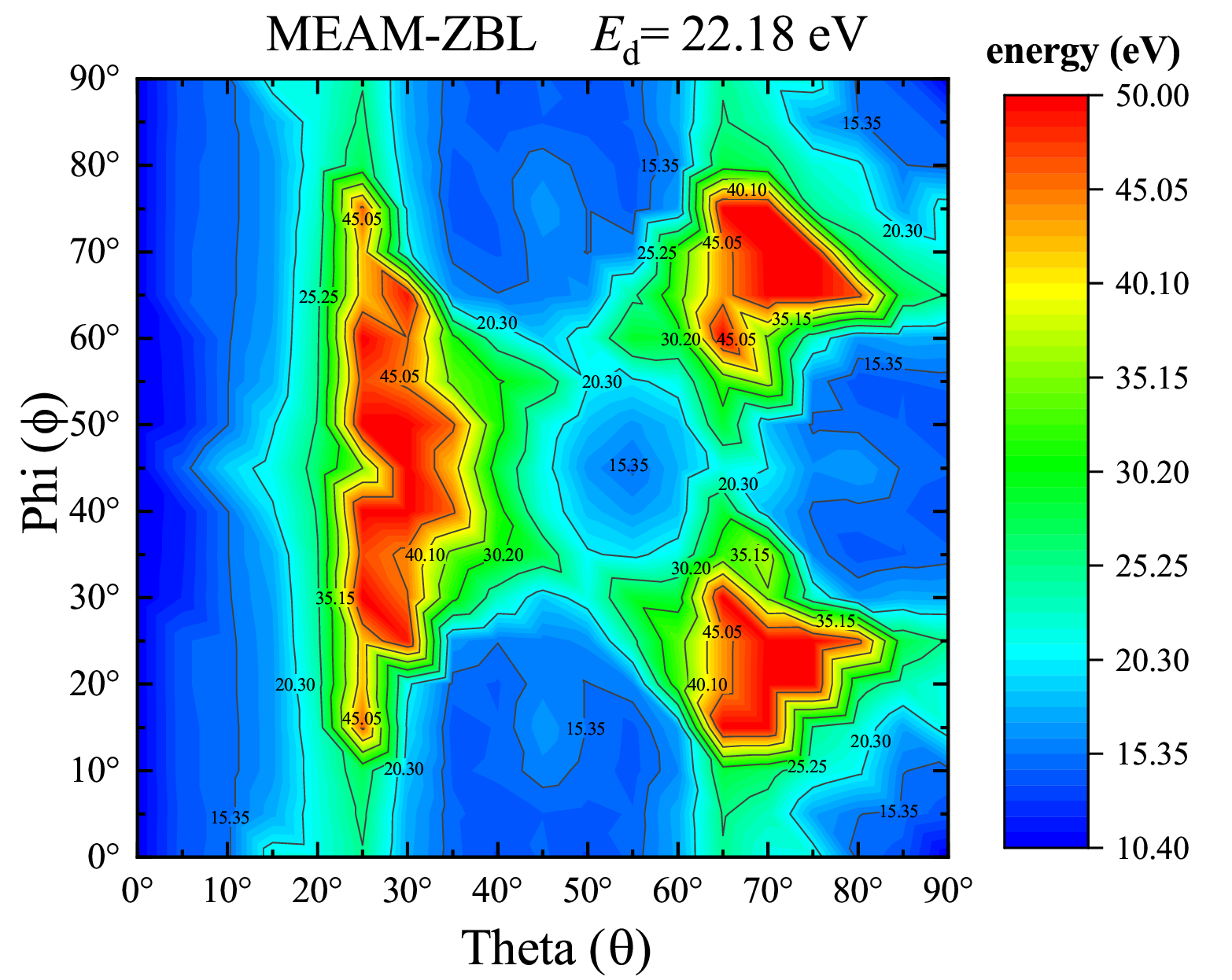}}
\subfloat[]{
\includegraphics[width=0.5\textwidth]{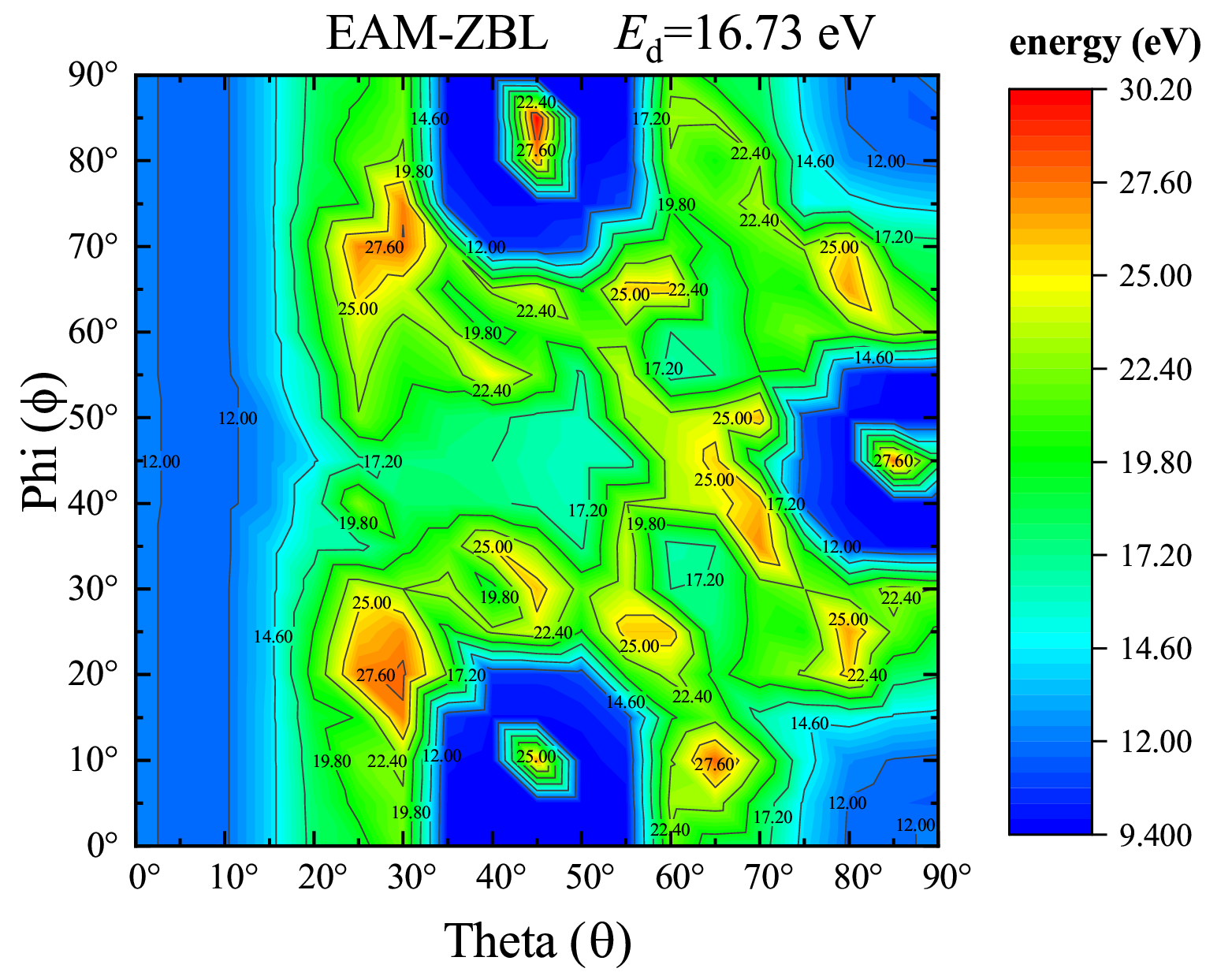}}

\caption{The corresponding $E_{\rm d}(\theta, \phi)$ calculated using
the (a) DP-ZBL, (b) MEAM-ZBL, and (c) MEAM-ZBL potentials. All the three potentials show similar dependence of $E_{\rm d}$ on the PKA initial direction ($\theta, \phi$)}
\end{figure}

\subsection{The evolution of collision cascade}

Here we provided the evolution of collision cascades induced by 1, 3, 4, 5 keV PKAs. All the three potentials including DP-ZBL, MEAM-ZBL, and EAM-ZBL can reproduce the reasonable trendy of displaced atoms $N_{\rm d}$ during the cascade simulations. Note that, the MEAM-ZBL potential has higher FP peaks than the EAM-ZBL and DP-ZBL potentials, because the MEAM-ZBL potential has overestimated the repulsive interaction below 1.5 \AA \ as demonstrated in FIG. \ref{fig:s3}. In summary, the DP-ZBL potential is sufficient to describe the evolution of FPs during the cascades events as well as the classical EAM-ZBL and MEAM-ZBL potentials.

\begin{figure}
	\centering
	\subfloat[]{
		\includegraphics[width=0.48\textwidth]{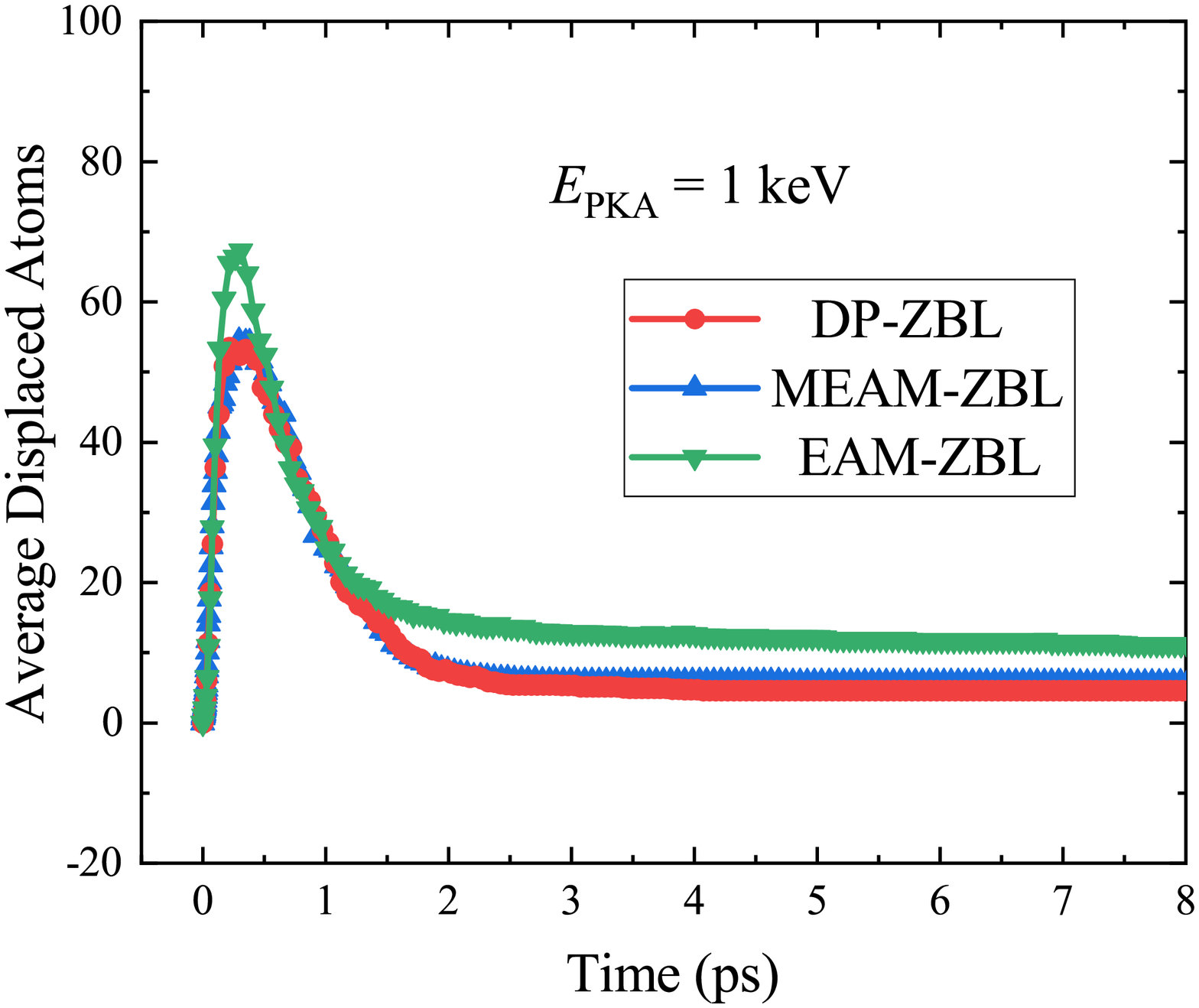}}
	\quad
	\subfloat[]{
		\includegraphics[width=0.48\textwidth]{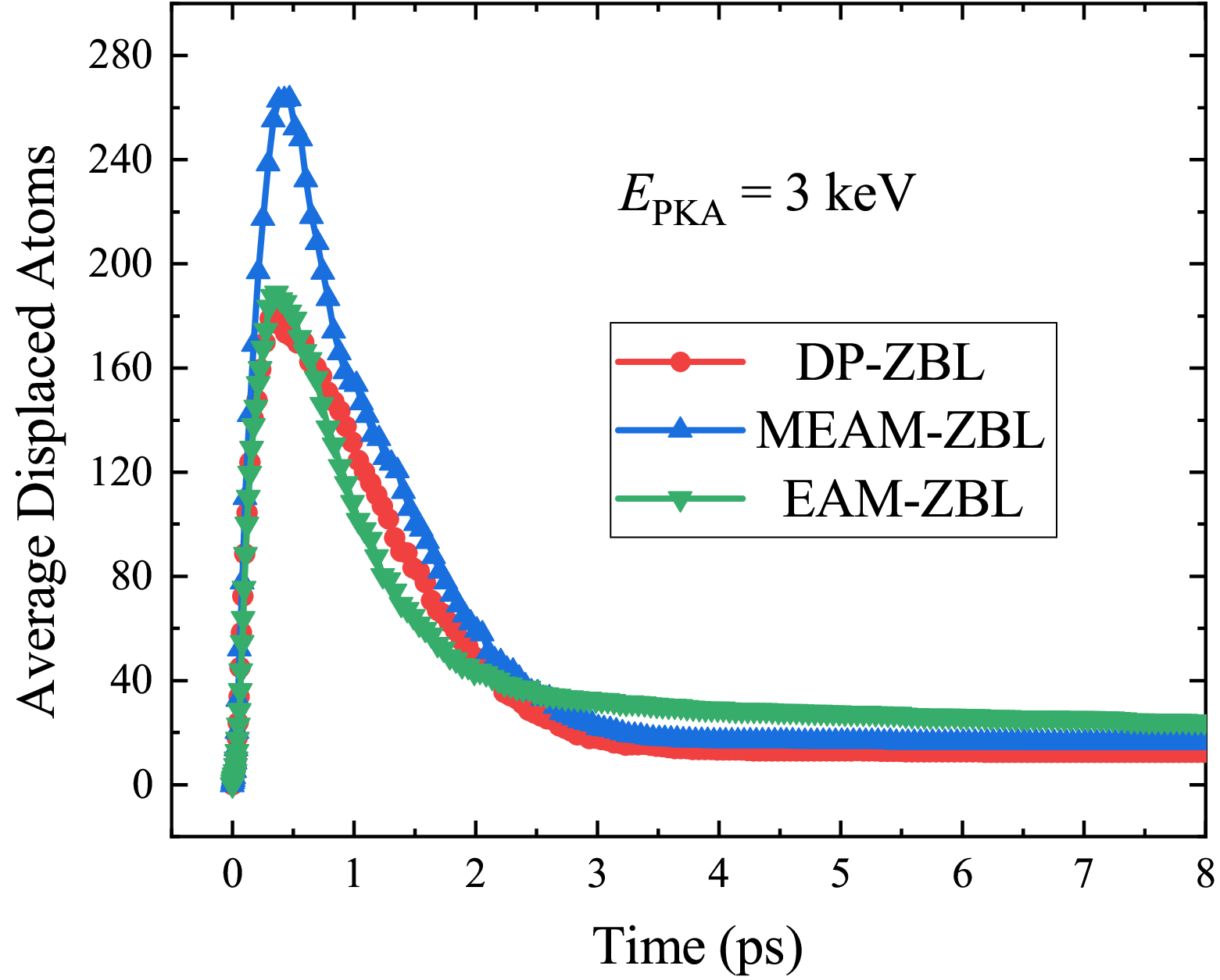}}
	\quad
	\subfloat[]{
		\includegraphics[width=0.48\textwidth]{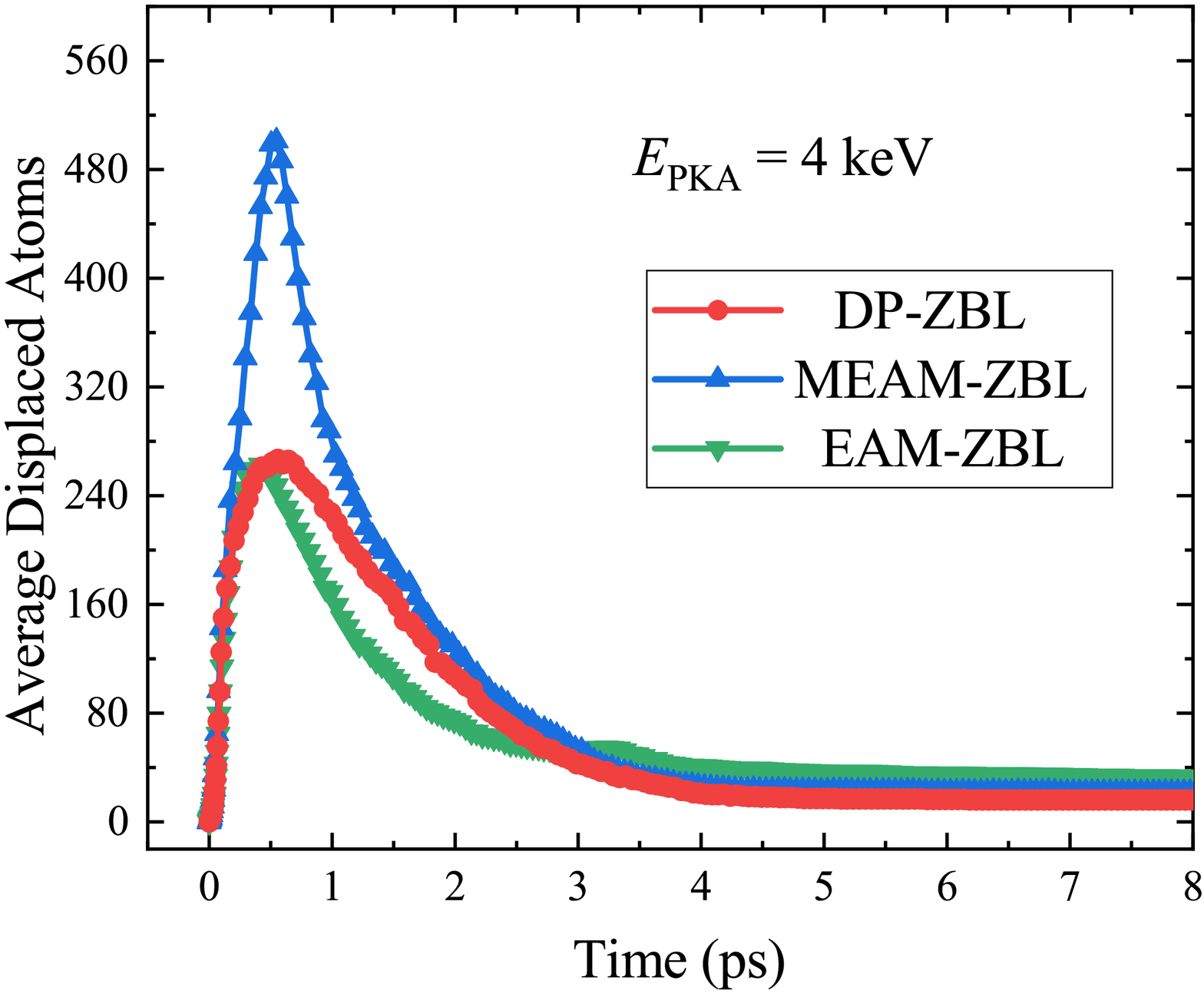}}
	\subfloat[]{
		\includegraphics[width=0.48\textwidth]{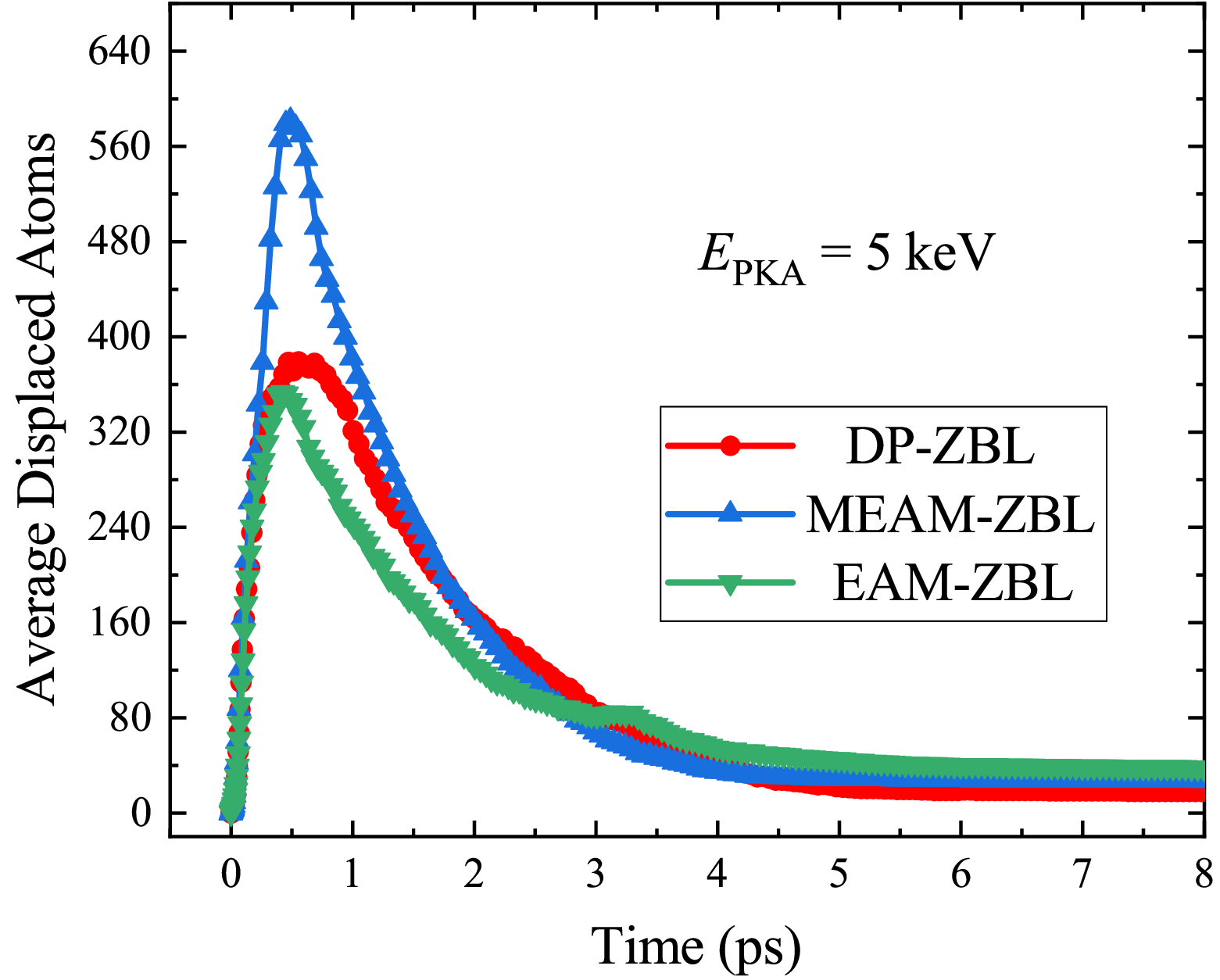}}
	
	\caption{The evolution of average displaced atoms number ($N_{\rm d}$) during the cascades events induced by 1 keV(a), 3 keV (b), 4 keV (c) and 5 keV (d) PKAs. Each point is the average of 10 independent simulations.}
\end{figure}

\newpage